# The Unique Origin of Colors of Armchair Carbon Nanotubes

Erik H. Hároz,[1,2] Benjamin Y. Lu,[1,2†] Pavel Nikolaev,[3] Sivaram Arepalli,[3] Robert H. Hauge,[2,4] and Junichiro Kono[1,2,5*]

[1]Department of Electrical and Computer Engineering, Rice University, Houston, Texas 77005, USA, [2]The Richard E. Smalley Institute for Nanoscale Science and Technology, Rice University, Houston, Texas 77005, USA, [3]Department of Energy Science, Sungkyunkwan University, Suwon 440-746, Korea, [4]Department of Chemistry, Rice University, Houston, Texas 77005, USA, [5]Department of Physics and Astronomy, Rice University, Houston, Texas 77005, USA.

**ABSTRACT:** The colors of suspended metallic colloidal particles are determined by their size-dependent plasma resonance, while those of semiconducting colloidal particles are determined by their size-dependent band gap. Here, we present a novel case for armchair carbon nanotubes, suspended in aqueous medium, for which the color depends on their size-dependent excitonic resonance, even though the individual particles are metallic. We observe distinct colors of a series of armchair-enriched nanotube suspensions, highlighting the unique coloration mechanism of these one-dimensional metals.

The size-dependent colors of suspended colloidal particles have fascinated researchers, engineers, and artists for centuries. While quantum confinement always plays a fundamental role, the coloration mechanism can differ depending on whether the particles are metallic or semiconducting. For metallic nanoparticles, their colors are determined by the free-carrier plasma resonance whose frequency depends on the electron density as well as the particle size and shape.[1] For semiconducting nanoparticles, the key parameter is the size-dependent fundamental band gap, i.e., the separation between the top of the valence band (HOMO) and the bottom of the conduction band (LUMO), which sensitively changes with quantum confinement, i.e., size.[2]

Here, we present a novel case for armchair single-walled carbon nanotubes (SWCNTs), suspended in aqueous medium, for which the origin of their color depends on the interband excitonic resonance even though the individual particles are gapless, i.e., metallic. Armchair nanotubes enjoy a rather special status among the SWCNT family. The structure of each member, or species, of the family is uniquely specified by a pair of integers, $(n,m)$, resulting in different species possessing different diameters, chiral angles, and electronic types (semiconducting or metallic).[3] Armchair SWCNTs are characterized by the simple relation $n = m$, i.e., $(n,n)$, and they are known to be the only truly gapless species with excellent electrical properties, exhibiting ballistic conduction even at room temperature.[4] At the same time, their one-dimensional characteristics combined with their linear band dispersions have attracted much fundamental interest for exploring many-body phenomena.[5] However, systematic studies of macroscopic ensembles of armchair nanotubes have been impossible due to the coexistence of different $(n,m)$ species of nanotubes in as-grown samples.

Recent years have seen impressive progress in post-growth separation of SWCNTs using a variety of methods. One of the most successful methods has been density gradient ultracentrifugation (DGU),[6-10] which can sort out different species of SWCNTs in bulk quantities according to their diameters, chiralities, and electronic types, enabling studies of $(n,m)$-dependent properties using standard macroscopic characterization measurements. In a recent report,[10] we provided unambiguous evidence of bulk enrichment of armchair nanotubes through DGU by utilizing wavelength-dependent resonant Raman scattering spectroscopy. We found that the Raman spectra were dominated by (6,6), (7,7), (8,8), (9,9), and (10,10) for samples enriched from carbon nanotubes synthesized by the high-pressure carbon monoxide (HiPco) method.

We studied the absorption properties of a series of armchair-enriched samples with different diameter distributions, exhibiting distinct colors (see Fig. 1). These samples (right of Fig. 1) were prepared through DGU with starting materials synthesized by CoMoCAT (average diameter, $d_{avg}$ = 0.83 nm), HiPco (batch no.189.2, $d_{avg}$ = 0.96 nm), HiPco (batch no.188.2, $d_{avg}$ = 1.1 nm), HiPco (batch no.107, $d_{avg}$ = 1.1 nm), laser ablation (NASA, $d_{avg}$ = 1.38 nm), and arc-discharge ($d_{avg}$ = 1.5 nm).

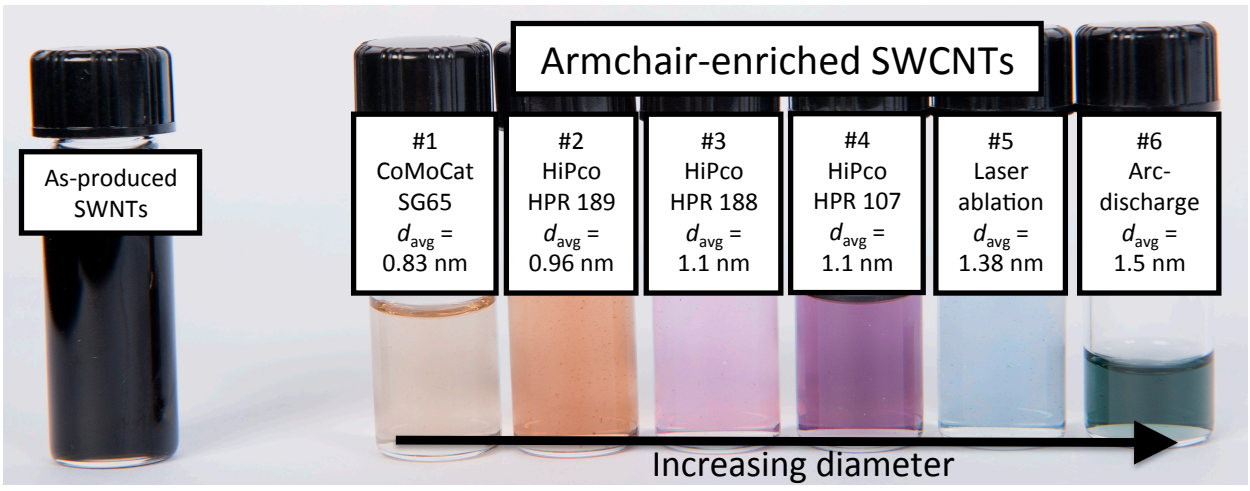

**Figure 1**. Pictures of armchair-enriched SWCNT suspensions. The black, "as-produced" vial on the left is typical of unsorted, SWCNT materials. On the right, various armchair-enriched samples with different diameters exhibit different, distinct colors.

On the left of Fig. 1 is a typical, as-produced SWCNT sample, which looks black because it contains a wide assortment of metallic and semiconducting SWCNTs with different diameters, absorbing everywhere in the visible optical range. Absorption spectra are shown in Fig. 2a for each armchair-enriched sample in Fig. 1, with (*n,m*)-peak assignments indicated. Figure 2b shows selected resonant Raman spectra, taken at specific excitation wavelengths for each sample, confirming the (*n,m*) assignments in Fig. 2a and highlighting the armchair-enriched nature of each sample (for more details on armchair enrichment, please see supplementary material, Fig. S2). The results of Fig. 1 and Fig. 2a-b are summarized in Table 1. In addition to demonstrating the intrinsic colors of each armchair species, these results demonstrate our ability to enrich armchair species of a particular diameter by careful choice of the starting SWCNT material.

| Sample # | SWCNT Material | Color | Main (*n,m*) Species |
|---|---|---|---|
| 1 | CoMoCAT SG65 | Yellow | (5,5), (6,6) |
| 2 | HiPco HPR 189.2 | Orange | (5,5), (6,6), (7,7), (8,8), (9,9) |
| 3 | HiPco HPR 188.2 | Magenta | (7,7), (8,8), (9,9) |
| 4 | HiPco HPR 107 | Purple | (8,8), (9,9) |
| 5 | Laser ablation JSC-385 | Cyan | (9,9), (10,10), (11,11) |
| 6 | Arc discharge P2 | Dark Green | (11,11), (12,12) |

**Table 1**. Correlation of SWCNT starting material with armchair-enriched suspension color and (*n,m*) composition.

Figure 2c shows a fit to the optical absorption spectrum of armchair-enriched sample #2, which was derived from HiPco SWCNT material. The fitting function consisted of a sum of six Lorentzian peaks, one for each armchair (10,10) through (5,5), along with a polynomial function used to fit the absorption baseline/background. While the absorption background has multiple possible intrinsic and extrinsic origins,[11,12] it is evident that the fit produces excellent agreement with the experimental data (adjusted $R^2$ goodness-of-fit value = 0.996). Similar Lorentizian fits for the absorption spectra of samples #1, 3, 4, 5 and 6 result in adjusted $R^2$ goodness-of-fit values = 0.984, 0.998, 0.997, 0.998 and 0.998, respectively (see supplementary Fig. S2 for more fitting examples). Of particular note is the highly symmetric lineshape of each peak, represented mathematically by the Lorentzian fitting function. Physically, each of the absorption features corresponds to an interband transition of a particular armchair species whose energy position varies roughly with inverse tube diameter.[5] The implication of this symmetric lineshape is that interband transitions in armchair SWCNTs are excitonic, contrary to expectations from their metallic character. Band-to-band optical transitions in SWCNTs should have an asymmetric absorption lineshape due to the one-dimensional van Hove singularities in the density of states. However, the one-dimensionality of armchair SWCNTs sufficiently reduces screening between electron and hole carriers to allow the formation of stable excitons, resulting in the observed symmetric lineshape. This agrees well with a recent theoretical study[13] as well as single-tube absorption measurements on a large-diameter [(21,21)] armchair nanotube that suggested the stability of excitons in metallic nanotubes.[14] As an additional consequence of such excitonic interactions, the continuum of absorption above the excitonic resonance should be significantly suppressed, again due to one-dimensionality, resulting in a Sommerfeld factor, the absorption intensity ratio of an unbound exciton to the free electron-hole pair above the band edge, less than unity.[15,16] Experimentally, this leads to a spectrally concentrated, narrow absorption band for each armchair species.

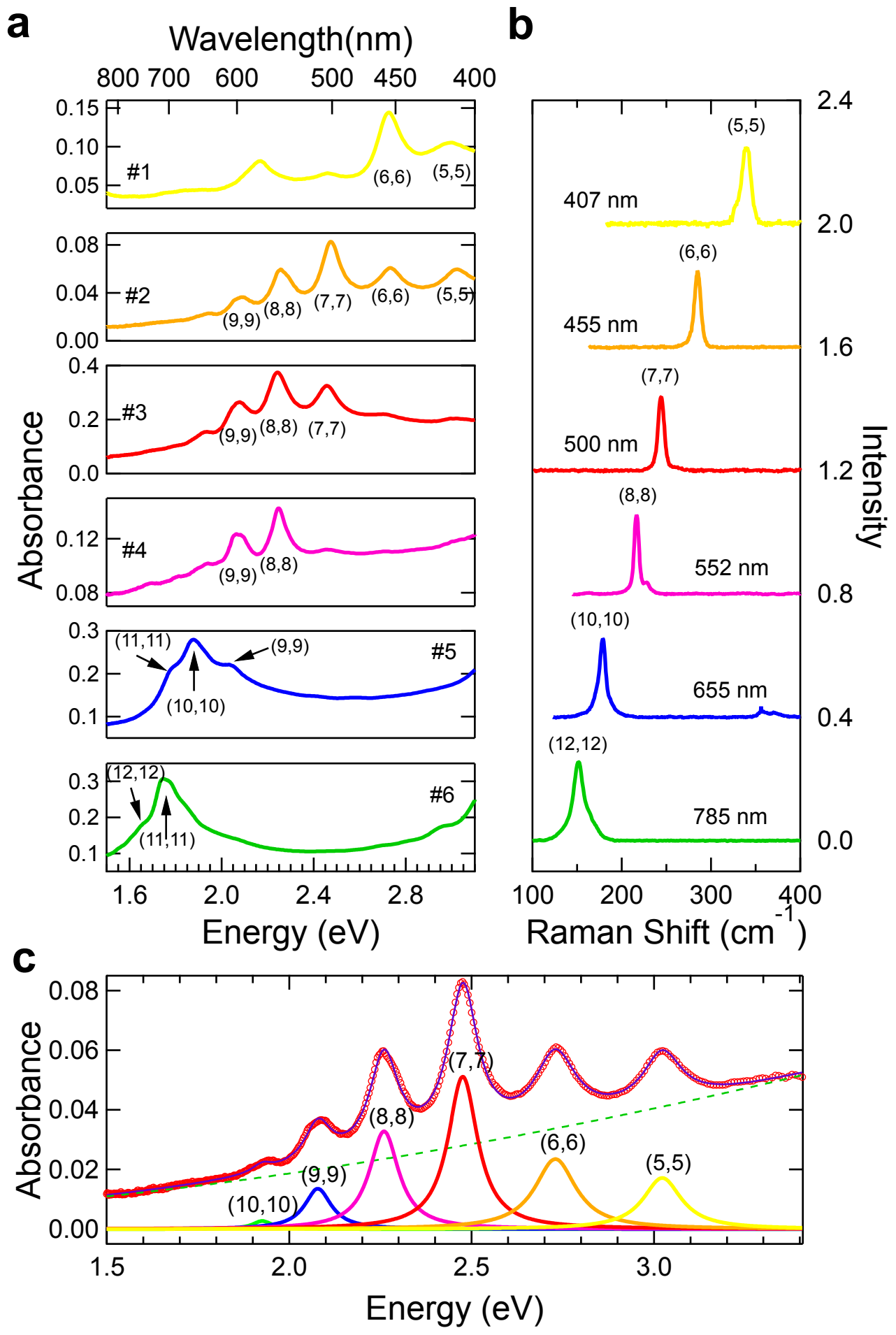

**Figure 2**. Optical spectra for armchair-enriched carbon nanotube samples with different diameters. **a**. Optical absorption spectra of armchair-enriched SWCNT suspensions, corresponding to the samples identified in Fig. 1. The main absorption peaks move toward higher energy with decreasing SWCNT diameter. **b**. Associated single-line resonant Raman radial breathing mode spectra for each enriched sample, further confirming the armchair enrichment of these samples and the peak assignments in absorption. **c**. Resulting excellent fit (thin purple line) of the optical absorption spectrum of armchair-enriched sample #2 (HiPco HPR 189.2, red open circles) to a sum of six Lorentzian peaks (thick, multi-color peaks), one for each armchair (10,10) through (5,5), on top of an absorption baseline (dashed green line).

Although armchair nanotubes are metallic in character, the colors of their suspensions are not determined by their plasmonic properties as in metallic nanoparticles, by their band gaps (which are zero) as in semiconducting nanoparticles, nor by their nanoparticle aggregate size (see supplementary Fig. S3). Rather, they are determined by a unique combination of band structure and selection rules for optical transitions. For armchair SWCNTs, optical transitions between the linear bands are forbidden;[17] the minimum energy required for absorption is the separation between the first van Hove singularities, which is *not* the HOMO-LUMO separation (see Fig. 3, bottom). Near the resonant absorption energy, the optical transition is excitonic with a strongly suppressed continuum above the band edge, resulting in sharp absorption only in the vicinity of the excitonic optical transition, as discussed above. This strong and narrow absorption peak for each armchair explains the apparent colors of our armchair-enriched suspensions when viewed in the context of subtractive color theory.

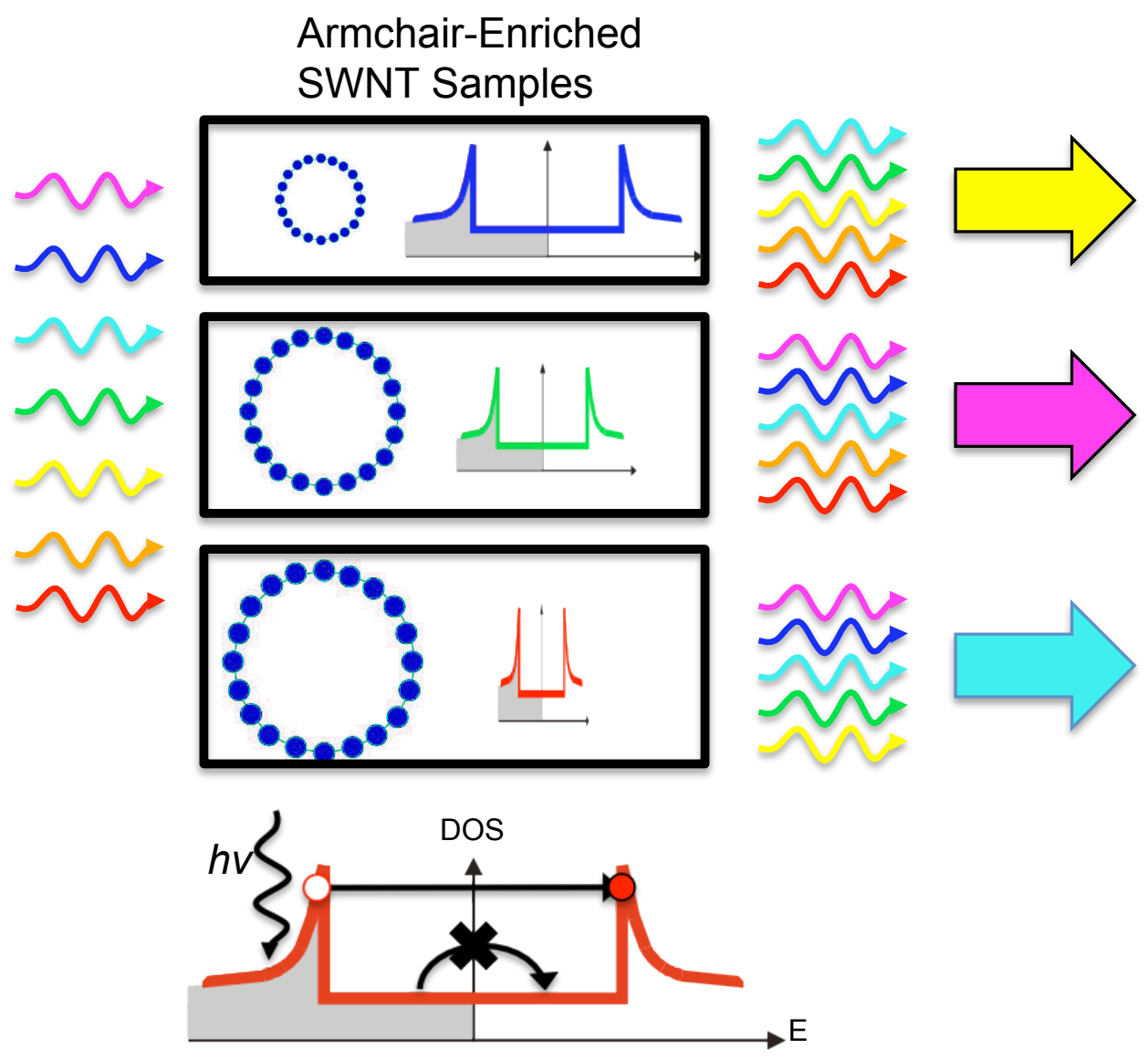

**Figure 3**. Subtractive coloration in armchair nanotubes due to interband excitonic resonance. Bottom: representative density-of-states for armchair SWCNTs with the lowest-energy allowed interband transition indicated. Top: as white light is passed through an armchair-enriched sample, photons with energies below the allowed interband transition are transmitted, while photons with energies resonant with exciton resonance associated with the allowed interband transition are absorbed; photons with energies larger than the exciton resonance are also weakly absorbed (see text). Since the energy of the exciton resonance is diameter-dependent, different armchair species have different, distinctive colors.

In subtractive color theory, the apparent color of a material is the result of certain wavelengths of visible light being removed or subtracted from white light as it is transmitted/reflected through/off of the material (illustrated in the upper panel of Fig. 3).[18] White light that has red (green; blue) light removed appears cyan (magneta; yellow) in color, respectively. In our armchair-enriched sample derived from CoMoCAT SWCNT material, sample #1 (top panel of Fig. 3), (5,5) and (6,6) armchairs absorb strongly in the blue, resulting in a yellow-colored suspension. Armchair-enriched laser ablation SWCNT material absorbs strongly in the red due to (9,9) and (10,10), producing a cyan-colored suspension. Armchair-enriched samples #3 and #4, which are produced from HiPco material, absorb mainly in the green region, producing suspensions that are colored magenta to purple, depending on the varying amounts of (7,7), (8,8), and (9,9) absorbed. Finally, armchair-enriched arc-discharge material (sample #6) appears green in color due to dual absorption in both the red, due to the first optical transitions of (10,10) and (11,11), and in the blue, due to the second optical transitions of (11,11) and (12,12). Hence, all observed colors for these suspensions obey the rules of subtractive color theory and are a result of the highly spectrally-concentrated absorption bands of armchair SWCNTs, a direct and macroscopic consequence of their excitonic optical properties.

In conclusion, we have shown that we can create a broad assortment of armchair-enriched carbon nanotube aqueous suspensions, produced by the density gradient ultracentrifugation method, from any number of SWCNT syntheses. Subsequently, we measured the optical properties of such suspensions through linear absorption spectroscopy and used lineshape fitting to confirm previous results that excitons do seem to exist in these one-dimensional metals. Finally, using such information in combination with subtractive color theory, we have established the origin of their colors as stemming from the narrow and symmetric absorption bands of armchair SWCNTs, a result of their unique excitonic properties and not the expected plasmon resonance observed in most metals. Such knowledge about these unusual low-dimensional metals will surely lead to further studies of other exotic phenomena in spectroscopy and materials science.

## ASSOCIATED CONTENT

**Supporting Information**. Methods: Sample preparation and Optical Measurements. Table S1: Table of density gradient parameters for preparation of samples. Figure S1: Raman excitation contour plots of sample #3. Figure S2: Additional examples of absorption spectra fitting for samples #3 & 4. Figure S3: Comparison between absorption spectra of sample #3 in aqueous suspension and thin-film forms. This material is available free of charge via the Internet at http://pubs.acs.org.

## AUTHOR INFORMATION

### Corresponding Author

* To whom correspondence should be addressed. E-mail: kono@rice.edu

### Present Addresses

†School of Medicine, New York University, New York, NY, USA

### Author Contributions

The manuscript was written through contributions of all authors. / All authors have given approval to the final version of the manuscript.

## ACKNOWLEDGMENT

The authors thank Juan G. Duque and Stephen K. Doorn of Los Alamos National Laboratory for providing some of Raman spectra presented here. The authors also thank Budhadipta Dan for providing the SEM image in supplementary material. This work was supported by the DOE/BES (DEFG02-06ER46308), the Robert A. Welch Foundation (C-1509), the Air Force Research La-boratories (FA8650-05-D-5807), NASA (NNJ05HI05C) and the World Class University Program at SKKU (R31-2008-000-10029-0).

## SYNOPSIS TOC

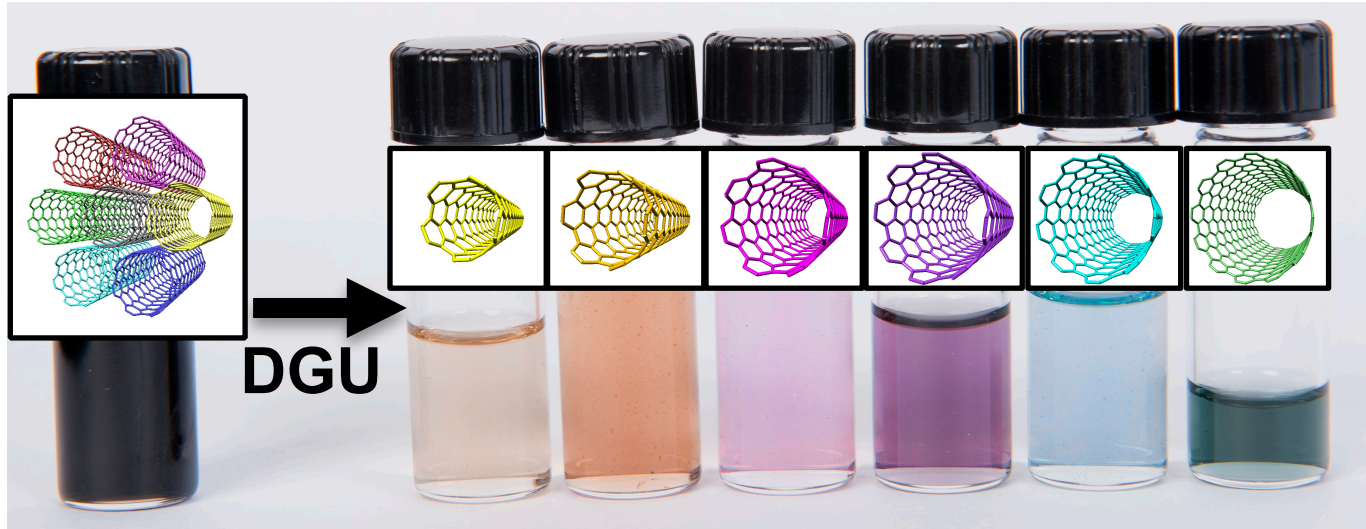